\documentclass[aps,prd,eqsecnum,preprint,tightenlines,nofootinbib]{revtex4-2}
\usepackage{graphicx,latexsym}

\def\bea{\begin{eqnarray}}
\def\eea{\end{eqnarray}}
\def\be{\begin{equation}}
\def\ee{\end{equation}}
\newcommand{\ub}[1]{\underline{#1}}
\newcommand{\ob}[1]{\overline{#1}}
\newcommand{\Pminus}{{\cal P}^-}
\newcommand{\Pplus}{{\cal P}^+}

\newcommand{\vaca}{|0\rangle_a}
\newcommand{\vacf}{|0\rangle_f}
\def\senk#1{\vec{#1}_\perp}

\begin{document}

\title{Effective potential between static sources \\
in quenched light-front Yukawa theory
}
\author{A.P.~Bray}
\author{S.S.~Chabysheva}
\affiliation{Department of Physics, University of Idaho, Moscow, Idaho 83844 USA}
\author{J.R.~Hiller}
\affiliation{Department of Physics, University of Idaho, Moscow, Idaho 83844 USA}
\affiliation{Department of Physics and Astronomy,
University of Minnesota-Duluth,
Duluth, Minnesota 55812 USA}

\date{\today}

\begin{abstract}

We compute a nonperturbative effective potential between two static fermions in
light-front Yukawa theory as a Hamiltonian eigenvalue problem.  
Fermion pair production is suppressed, to make 
possible an exact analytic solution in the form of a coherent state of
bosons that form clouds around the sources.  The effective potential is
essentially an interference term between individual clouds.  The model
is regulated with Pauli-Villars bosons and fermions, to achieve
consistent quantization and renormalization of masses and couplings.
This extends earlier work on scalar Yukawa theory where Pauli-Villars
regularization did not play a central role.  The key result is that
the nonperturbative solution restores rotational symmetry even though
the light-front formulation of Yukawa theory, with its preferred axis,
appears antithetical to such a symmetry.

\end{abstract}

\maketitle

\section{Introduction}  \label{sec:Introduction}

The current understanding of quantum chromodynamics (QCD) as the
theory of the strong interactions is that it provides for
the confinement of quarks and gluons.  Calculations within
lattice QCD~\cite{Gattringer,Rothe} have confirmed this~\cite{NeccoSommer,Bazavov,Brambilla}
as have analytic calculations~\cite{Karbstein,Lee,Giusti},\footnote{For earlier
discussion of static potentials in QCD, see~\protect\cite{Appelquist,Brown}.}
not to mention the experimental evidence for the absence of free quarks and gluons.
What has not been established is the derivation of
confinement within a light-front quantization of the 
theory~\cite{LFreview1,LFreview2,LFreview3,LFreview4,WhitePaper,LFreview5}.
This is something that must be nonperturbative or at least
an all-orders resummation of perturbation theory.

We wish to explore how such a nonperturbative light-front 
calculation might be done.  Here we extend earlier 
work~\cite{ScalarYukawa}\footnote{In \protect\cite{ScalarYukawa} there are 
two typographical errors.  The factor $g/m$ in Eq.~(3.27) should
be $g/(2m)$, and the $(q^+)^2$ in the numerator of 
Eq.~(3.29) should be $q^+$.}
on quenched scalar Yukawa theory to include fermion sources.
The basic approach is to consider the effective potential
for two static sources as a function of their separation,
computed from the change in the eigenenergy of the system
relative to the energy of two well-separated sources.
This involves renormalization of the source mass and, as
we will show, renormalization of the coupling to bosons.
Regularization is provided by the inclusion of Pauli-Villars (PV)
fermions and bosons.\footnote{For earlier uses of PV regularization
in nonperturbative light-front calculations, see \cite{Yukawa}
for Yukawa theory 
and \cite{QED,ChiralLimit,TwoPhotons,VacPol,ArbGauge} for quantum electrodynamics.}
The PV fermions provide a convenient
simplification of the light-front quantization by eliminating
what are known as instantaneous fermion terms from the 
light-front Hamiltonian; this is what makes the analytic
solution possible.  The PV bosons regulate the 
self-energy corrections to the fermion mass.

Some will be concerned that a method based on PV regularization
cannot be extended to QCD, there being a general prejudice
against PV regularization of nonabelian gauge theories.
However, as shown in \cite{LFreview5}, a consistent PV
regularization can be constructed.  The key is that the 
definition of the gauge transformation must be extended
when the Lagrangian is extended to include the PV fields.
In other words, there does remain a gauge invariance of the QCD 
Lagrangian when PV fields are included.  In addition,
there is a BRST symmetry when the gauge is fixed
covariantly.

Our definition of light-front coordinates~\cite{Dirac} is to take
$x^\pm=t\pm z$, with $x^+$ as the light-front time,
and $\vec{x}_\perp=(x,y)$ in the transverse.  The 
conjugate light-front energy is $p^-=E-p_z$, and the 
light-front momentum is
$\ub{p}=(p^+\equiv E+p_z,\vec{p}_\perp\equiv(p_x,p_y))$.
The mass-shell condition $p^2=m^2$ becomes
$p^-=(\vec{p}_\perp^{\,2}+m^2)/p^+$.
Further details can be found in the review~\cite{LFreview5}.

For static sources, fixed in a lab frame, the eigenstates
are no longer eigenstates of light-front energy or 
momentum.\footnote{For earlier work with static sources
on the light-front, see \protect\cite{Thorn,Klindworth,transverselattice,Blunden}.}
The ordinary momentum is zero, including the $z$ component.
We therefore seek eigenstates of the ordinary energy with
use of the operator ${\cal E}\equiv\frac12(\Pminus+\Pplus)$
but do so in terms of light-front coordinates.
Such a choice is also motivated by the fact that the definition
of an effective potential is in the dependence of the ordinary
energy on the source separation.  Most calculations
in light-front-quantized theories need not make this
distinction because they use a basis where the light-front
momentum is held fixed.

The parameters of the Yukawa Lagrangian are renormalized
by fixing the mass of the dressed fermion state at a 
physical value $m$ and by requiring the effective 
potential to be of the standard Yukawa form $-\frac{g^2}{4\pi R}e^{-\mu R}$,
where $R$ is the source separation, $\mu$ the renormalized boson mass,
and $g$ the physical coupling.  For the quenched case
considered here, the boson mass in the Lagrangian is
not actually renormalized.  The form of the effective
potential is obtained, including its rotational symmetry.

In Sec.~\ref{sec:Yukawa} we summarize the structure of
light-front Yukawa theory with PV regularization and 
of the Hamiltonian eigenvalue problem for static
fermions.  The solution for a single source is
developed in Sec.~\ref{sec:Single} and for a double
source in Sec.~\ref{sec:Double}.  We summarize the
results in Sec.~\ref{sec:Summary} and leave some details
to Appendices.

\section{Light-front Yukawa theory}  \label{sec:Yukawa}

The Lagrangian for PV-regulated Yukawa theory 
is~\cite{LFreview5}\footnote{In Eq.~(223) of \protect\cite{LFreview5} the 
Yukawa Lagrangian is expressed explicitly in terms of separate fields
rather than a more general sum.  Here instead we present it
in a form analogous to the QED Lagrangian in Eq.~(131).}
\be
{\cal L}=\sum_k r_k\left[\frac12(\partial_\mu\phi_k)^2-\frac12\mu_k^2\phi_k^2\right]
+\sum_i s_i \ob{\psi}_i(i\gamma^\mu\partial_\mu-m_i)\psi_i
-g_0\sum_{ijk}\beta_i\beta_j\xi_k\ob{\psi}_i\psi_j\phi_k,
\ee
where $k=i=j=0$ correspond to physical fields and positive
integers to PV fields.  The $r_k$ and $s_i$ are metric signatures,
with $r_0=s_0=1$ for the physical fields, $r_1=s_1=-1$ for
the first PV fields, and the remainder (if any) determined
by the constraints necessary for a consistent theory.  The
factors of $\beta_i$ and $\xi_k$ provide for adjustments of
the relative couplings of the PV fields, with $\beta_0=\xi_0=1$ as a definition
of $g_0$ as the bare coupling for physical fields. The
regularization of loops is provided by the constraints~\cite{LFreview5}
\be  \label{eq:constraints}
\sum_k r_k \xi_k^2=0\;\;\mbox{and}\;\; \sum_i s_i\beta_i^2=0,
\ee
because one or the other of these combinations appears
for each line in a loop when summed over the PV contributions.
The $r_k$ and $s_i$ factors come from
the propagator, and the $\xi_k$ and $\beta_i$ come from the 
vertices at the ends of the line.  The leading UV divergence
then involves these sums and is canceled by the constraints (\ref{eq:constraints}).
For example, a boson line carrying momentum $q$ contributes
$\sum_k r_k \xi_k^2/q_\perp^2 + {\cal O}\left(\sum_k r_k m_k ^2\xi_k^2/q_\perp^4\right)$,
and the leading term is zero.
This cancellation will be seen explicitly in the next section.
The nonperturbative calculation is effectively a resummation
of the regulated loop expansion.

The mode expansions for the boson fields are
\be
\phi_k(x)=\int\frac{d\ub{p}}{\sqrt{16\pi^3 p^+}}
        \left[a_k(\ub{p})e^{-ip\cdot x}+a_k^\dagger(\ub{p})e^{ip\cdot x}\right],
\ee
with the nonzero commutation relations being
\be \label{eq:scalarcommutators}
{[}a_k(\ub{p}),a_{k'}^\dagger(\ub{p'})]=\delta(\ub{p}-\ub{p'})
             \equiv r_k\delta_{kk'}\delta(p^+-p^{\prime +})\delta(\senk{p}-\senk{p'}).
\ee
The factor $r_k=\pm1$ fixes the sign of the metric for the field.

The fermion field satisfies the Euler--Lagrange equation
\be
s_i(i\gamma^\mu\partial_\mu-m_i)\psi_i-g_0\sum_{jk}\beta_i\beta_j\xi_k\phi_k\psi_j=0.
\ee
To separate the dynamical part $\psi_{i+}\equiv\frac12\gamma^0\gamma^+\psi_i$ from the
constrained part $\psi_{i-}\equiv\frac12\gamma^0\gamma^-\psi_i$, we project this
Euler-Lagrange equation using $\frac12\gamma^0\gamma^\pm$ to yield the following two equations:
\be
s_i i\partial_+\psi_{i+}-s_i(-i\vec{\alpha}_\perp\cdot\partial_\perp+\beta m_i)\psi_{i-}
  -g_0\sum_{jk}\beta_i\beta_j\xi_k\phi_k\psi_{j-}=0
\ee
and
\be \label{eq:constrainteq}
s_i i\partial_-\psi_{i-}-s_i(-i\vec{\alpha}_\perp\cdot\partial_\perp+\beta m_i)\psi_{i+}
  -g_0\sum_{jk}\beta_i\beta_j\xi_k\phi_k\psi_{j+}=0,
\ee
with $\gamma^\mu=(\beta,\beta\vec{\alpha})$.
Multiplication of the constraint equation (\ref{eq:constrainteq})
by $s_i\beta_i$ and a sum over $i$ eliminates
the interaction,\footnote{In \protect\cite{LFreview5}, the analogous process for QED contains
an error in the line above Eq.~(138).  The factors $(-1)^i\sqrt{\beta_i}$ should be 
replaced with $s_i\beta_i$.}
such that the constrained part of the summed fermion field $\psi\equiv\sum_i\beta_i\psi_i$
satisfies
\be
i\partial_-\psi_--(-i\vec{\alpha}_\perp\cdot\partial_\perp\psi_++\beta \sum_i\beta_i m_i\psi_{i+})=0.
\ee
This is just the constraint equation for a free fermion.  We can then construct the
Hamiltonian from the free-fermion mode expansion
\be
\psi_i(x)=\int\frac{d\ub{p}}{\sqrt{16\pi^3 p^+}}\sum_{s=\pm1/2}
     \left[b_{is}(\ub{p})u_{is}(\ub{p})e^{-ip\cdot x}
           +d^\dagger_{is}(\ub{p})v_{is}(\ub{p}) e^{ip\cdot x}\right],
\ee
with
\be
u_{is}(\ub{p})\equiv\frac{1}{\sqrt{p^+}}\left[p^++\senk{\alpha}\cdot\senk{p}+\beta m_i\right]\chi_s, \;\;
v_{is}(\ub{p})\equiv\frac{1}{\sqrt{p^+}}\left[p^++\senk{\alpha}\cdot\senk{p}-\beta m_i\right]\chi_{-s}
\ee
and
\be
\chi_{+\frac12}=\frac{1}{\sqrt2}\left(\begin{array}{c} 1\\0\\1\\0 \end{array}\right), \;\;
\chi_{-\frac12}=\frac{1}{\sqrt2}\left(\begin{array}{c} 0\\1\\0\\-1 \end{array}\right).
\ee 
The nonzero anticommutators are
\be
\{b_{is}(\ub{p}),b_{js'}^\dagger(\ub{p}')\}=s_i\delta_{ij}\delta_{ss'}\delta(\ub{p}-\ub{p'}), \;\;
\{d_{is}(\ub{p}),d_{js'}^\dagger(\ub{p}')\}=s_i\delta_{ij}\delta_{ss'}\delta(\ub{p}-\ub{p'}),
\ee
and $s_i=\pm1$ sets the metric.

Instantaneous fermion interactions do not appear.  However, the physics of
these interactions has not been lost; they
are present implicitly and restored explicitly in the limit of infinite PV masses,
which shrinks a PV-fermion exchange to a contact interaction~\cite{TwoPhotons}.

The light-front Hamiltonian density is then
\bea
{\cal H}&=&\sum_k r_k\left[\frac12(\senk{\partial}\phi_k)^2+\frac12\mu_k^2\phi_k^2\right]
  +\sum_i s_i\left[\psi^\dagger_{i+}(i\senk{\alpha}\cdot\senk{\partial}-\beta m_i)
       \frac{1}{i\partial_-}(i\senk{\alpha}\cdot\senk{\partial}-\beta m_i)\psi_{i+}\right] \nonumber \\
    &&   +g_0\sum_{ijk}\beta_i\beta_j\xi_k\phi_k\ob{\psi}_i\psi_j.
\eea
The Hamiltonian 
\be
\Pminus\equiv \int d\ub{x}\left.:{\cal H}:\right|_{x^+=0}
  =\Pminus_{0a}+\Pminus_{0f}+\Pminus_{\rm n.p.}+\Pminus_{\rm pair}
\ee
is specified by
\be
\Pminus_{0a}=\sum_k r_k \int d\ub{q}\frac{\mu_k^2+q_\perp^2}{q^+}a_k^\dagger(\ub{q})a_k(\ub{q}),
\ee
\be
\Pminus_{0f}=\sum_i s_i \int  d\ub{p}\frac{m_i^2+\senk{p}^2}{p^+}\sum_s
       \left[b_{is}^\dagger(\ub{p})b_{is}(\ub{p})+d_{is}^\dagger(\ub{p})d_{is}(\ub{p})\right],
\ee
\bea \label{eq:Pnp}
\Pminus_{\rm n.p.}&=&g_0\sum_{ijk}\beta_i\beta_j\xi_k\int d\ub{x}\int\frac{d\ub{q}}{\sqrt{16\pi^3 q^+}}
  \left[a_k(\ub{q})e^{-i\ub{q}\cdot\ub{x}}+a^\dagger_k(\ub{q})e^{i\ub{q}\cdot\ub{x}}\right] \\
  &&\times \int \frac{d\ub{p}_1 d\ub{p}_2}{16 \pi^3}\sum_s\left\{\left[
     \left(\frac{m_i}{p_1^+}+\frac{m_j}{p_2^+}\right)b^\dagger_{is}(\ub{p}_1)b_{js}(\ub{p}_2)\right.\right. \nonumber \\
  && \rule{1in}{0mm}+\left.\left(\frac{\sqrt{2}\vec{\epsilon}_{-2s}\cdot\vec{p}_{1\perp}}{p_1^+}
           +\frac{\sqrt{2}\vec{\epsilon}^{\,*}_{2s}\cdot\vec{p}_{2\perp}}{p_2^+}\right)b^\dagger_{is}(\ub{p}_1)b_{j,-s}(\ub{p}_2)
     \right]e^{i(\ub{p}_1-\ub{p}_2)\cdot\ub{x}}  \nonumber \\
  && \rule{1in}{0mm}+\left[\left(\frac{m_i}{p_1^+}+\frac{m_j}{p_2^+}\right)d^\dagger_{js}(\ub{p}_2)d_{is}(\ub{p}_1)\right. \nonumber \\
  && \rule{1in}{0mm}-\left.\left.\left(\frac{\sqrt{2}\vec{\epsilon}_{-2s}\cdot\vec{p}_{1\perp}}{p_1^+}
           +\frac{\sqrt{2}\vec{\epsilon}^{\,*}_{2s}\cdot\vec{p}_{2\perp}}{p_2^+}\right)d^\dagger_{js}(\ub{p}_2)d_{i,-s}(\ub{p}_1) \nonumber
     \right]e^{i(\ub{p}_2-\ub{p}_1)\cdot\ub{x}}
     \right\},
\eea
and
\bea
\Pminus_{\rm pair}&=&g_0\sum_{ijk}\beta_i\beta_j\xi_k\int d\ub{x}\int\frac{d\ub{q}}{\sqrt{16\pi^3 q^+}}
  \left[a_k(\ub{q})e^{-i\ub{q}\cdot\ub{x}}+a^\dagger_k(\ub{q})e^{i\ub{q}\cdot\ub{x}}\right] \\
  &&\times \int \frac{d\ub{p}_1 d\ub{p}_2}{16 \pi^3}\sum_s\left\{\left[
     \left(\frac{m_i}{p_1^+}-\frac{m_j}{p_2^+}\right)b^\dagger_{is}(\ub{p}_1)d^\dagger_{j,-s}(\ub{p}_2)\right.\right. \nonumber \\
  && \rule{1in}{0mm}+\left.\left(\frac{\sqrt{2}\vec{\epsilon}_{2s}\cdot\vec{p}_{1\perp}}{p_1^+}
           +\frac{\sqrt{2}\vec{\epsilon}^{\,*}_{-2s}\cdot\vec{p}_{2\perp}}{p_2^+}\right)b^\dagger_{is}(\ub{p}_1)d^\dagger_{j,s}(\ub{p}_2)
     \right]e^{i(\ub{p}_1+\ub{p}_2)\cdot\ub{x}}  \nonumber \\
  && \rule{1in}{0mm}+\left[\left(\frac{m_i}{p_1^+}-\frac{m_j}{p_2^+}\right)b_{js}(\ub{p}_2)d_{i,-s}(\ub{p}_1)\right. \nonumber \\
  && \rule{1in}{0mm}-\left.\left.\left(\frac{\sqrt{2}\vec{\epsilon}_{-2s}\cdot\vec{p}_{1\perp}}{p_1^+}
           +\frac{\sqrt{2}\vec{\epsilon}^{\,*}_{2s}\cdot\vec{p}_{2\perp}}{p_2^+}\right)b_{js}(\ub{p}_2)d_{i,s}(\ub{p}_1) \nonumber
     \right]e^{-i(\ub{p}_1+\ub{p}_2)\cdot\ub{x}}
     \right\},
\eea
with $\sqrt{2}\vec{\epsilon}_{2s}\equiv -(2s,i)$ a two-dimensional, transverse 
vector~\cite{McCartorRobertson}, where $s$ is the spin index and $i=\sqrt{-1}$.  
The integral over $\ub{x}$ has been left undone in the interaction terms, to accommodate
the source wave packets introduced in the next sections.  For the quenched theory,
the term $\Pminus_{\rm pair}$ is, of course, dropped.  Also, we will limit our
work to the fermion sector, and antifermion terms in $\Pminus_{0f}$ and $\Pminus_{\rm n.p.}$
will play no role.

The light-front momentum operator is $\Pplus=\Pplus_a+\Pplus_f$ with
\be
\Pplus_a=\sum_k r_k \int d\ub{q}\,q^+a_k^\dagger(\ub{q})a_k(\ub{q})
\ee
and
\be
\Pplus_f=\sum_i s_i \int d\ub{p}\, p^+\sum_s\left[b^\dagger_{is}(\ub{p})b_{is}(\ub{p})+d^\dagger_{is}(\ub{p})d_{is}(\ub{p})\right].
\ee
We can then define the ordinary energy operator
\be
{\cal E}=\frac12(\Pminus+\Pplus).
\ee
In the next sections we explore eigenstates of ${\cal E}$ that
are associated with one or two static fermion sources.

\section{Single static source}  \label{sec:Single}

We first consider a single source at $\pm\vec{R}/2$ to establish the
renormalization of the fermion mass.  The static fermion is
described by a wave packet centered at $\ub{p}=(m,\senk{0})$
in momentum space and at $\ub{x}=(\mp R_z,\pm\senk{R}/2)$ in 
coordinate space on the $x^+=0$ slice.\footnote{The translation 
of the longitudinal coordinate, between a lab-fixed frame and
light-front coordinates, is illustrated in Fig.~1 of 
\protect\cite{ScalarYukawa}.  We have $z=\pm R_z/2$ fixed
and $x^+=t+z=0$; therefore, the ordinary time is $t=\mp R_z/2$
and the light-front spatial coordinate is $x^-=t-z=\mp R_z$.}  
The static fermion state with spin $s$ and PV type $i$ is then
\be \label{eq:Fpm}
|F^\pm_{is}\rangle=\int_{p^+>0} d\ub{p}F^\pm(\ub{p})b^\dagger_{is}(\ub{p})\vacf,
\ee
with the function $F^\pm$ peaked at $\ub{p}=(m,\senk{0})$
and its Fourier transform $\psi^\pm(\ub{x})$ peaked at 
$(\mp R_z,\pm\senk{R}/2)$.  Here $\vacf$ is the vacuum annihilated
by $b_{is}$.  The physical and PV fermions are
all static at the same location and at the same light-front momentum.

The Fourier transform is defined as
\be
F^\pm(\ub{p})=\int\frac{d\ub{x}}{\sqrt{16\pi^3}}e^{i\ub{p}\cdot\ub{x}}\psi^\pm(\ub{x}).
\ee
For the inverse, where the $p^+$ integration is limited to positive values,
we take advantage of the narrow peak in $F^\pm$ to 
extend the $p^+$ integral to $-\infty$
\be \label{eq:psi}
\psi^\pm(\ub{x})=\int\frac{d\ub{p}}{\sqrt{16\pi^3}}e^{-i\ub{p}\cdot\ub{x}}F^\pm(\ub{p}).
\ee
For a static source at $\ub{x}=(\mp R_z,\pm\senk{R}/2)$ we require that
\be \label{eq:source}
|\psi^\pm(\ub{x})|^2\rightarrow \delta(x^-\pm R_z)\delta(\senk{x}\mp\senk{R}/2).
\ee
The common normalization
\be
1=\int d\ub{x} |\psi^\pm(\ub{x})|^2=\int d\ub{p} |F^\pm(\ub{p})|^2
\ee
is fixed by requiring the indefinite norm
\be
\langle F^\pm_{is}|F^\pm_{is}\rangle
=\int d\ub{p}' d\ub{p} F^{\pm*}(\ub{p}') F^\pm(\ub{p}) s_i \delta(\ub{p}'-\ub{p}) 
=s_i\int d\ub{p}\,|F^\pm(\ub{p})|^2=s_i.  \nonumber
\ee

From this static-fermion state, we build a fermion state dressed by a cloud of bosons
in a coherent state as the {\em ansatz} for the energy eigenstate:
\be \label{eq:GFs}
|G^\pm F^\pm; s\rangle\equiv \sum_i C^\pm_{i} |G^\pm_i\rangle|F^\pm_{is}\rangle
\ee
with
\be
|G^\pm_i\rangle\equiv \sqrt{Z^\pm_i} 
\left[\prod_k \exp\left(\int d\ub{q} G^\pm_{ki}(\ub{q})a^\dagger_k(\ub{q})\right)\right]\vaca.
\ee
and $\vaca$ the vacuum annihilated by $a_k$.
Because the spin-flip terms in $\Pminus_{\rm n.p.}$ are proportional to $\senk{p}$
and the static state requires $\langle\senk{p}\rangle$ to be zero, the eigenstate is diagonal in spin,
and the coefficient $C^\pm_i$ and the functions $G^\pm_{ki}$ are independent of $s$.
The $\sqrt{Z^\pm_i}$ are normalization factors for the coherent state, given by
\be
Z^\pm_i=\exp\left(-\sum_k r_k\int d\ub{q} |G^\pm_{ki}(\ub{q})|^2\right).
\ee
This is then to be the solution to 
\be
{\cal E}|G^\pm F^\pm; s\rangle=E^\pm|G^\pm F^\pm; s\rangle,
\ee
with $E^\pm=m$ for the ground state, which is the state of interest.
Each term in the sum over $i$ is also an eigenstate of the boson 
annihilation operators $a_k$, as is always the case for a coherent state:
\be  \label{eq:aG}
a_k(\ub{q})|G^\pm_i\rangle=r_k G^\pm_{ki}(\ub{q})|G^\pm_i\rangle.
\ee

We begin with a projection of the eigenvalue problem onto a static fermion of type $i$
\be
s_i\langle F^\pm_{is}|{\cal E}|G^\pm F^\pm; s\rangle=E^\pm s_i\langle F^\pm_{is}|G^\pm F^\pm; s\rangle
   =E^\pm C^\pm_i|G^\pm_i\rangle.
\ee
This can be reduced with use of the following projections for individual terms
in ${\cal E}=\frac12(\Pminus_{0f}+\Pplus_f+\Pminus_{0a}+\Pplus_a+\Pminus_{\rm n.p.})$:
\bea
s_i\langle F^\pm_{is}|\frac12(\Pminus_{0f}+\Pplus_f)|G^\pm F^\pm; s\rangle
&=&\frac12C^\pm_i\int d\ub{p} \left[\frac{m_i^2+p_\perp^2}{p^+}+p^+\right] |F_i^\pm(\ub{p})|^2|G^\pm_i\rangle \\
&=&C^\pm_i\left(\frac{m_i^2}{2m}+\frac{m}{2}\right)|G^\pm_i\rangle, \nonumber
\eea
\be
s_i\langle F^\pm_{is}|\frac12(\Pminus_{0a}+\Pplus_a)|G^\pm F^\pm; s\rangle
=\frac12 C^\pm_{i}  \sum_k  \int d\ub{q} \left[\frac{\mu_k^2+q_\perp^2}{q^+}+q^+\right] 
 a_k^\dagger(\ub{q})G^\pm_{ki}(\ub{q})|G^\pm_i\rangle,
\ee
and
\bea \label{eq:projectedPnp}
\lefteqn{s_i\langle F^\pm_{is}|\frac12\Pminus_{\rm n.p.}|G^\pm F^\pm; s\rangle
=\frac12 g_0 \beta_i \sum_{jk} s_j \beta_j\xi_k C^\pm_j \frac{m_i+m_j}{m}}&& \\
&&\times \int \frac{d\ub{q}}{\sqrt{16\pi^3 q^+}}\left[r_k G^\pm_{kj}(\ub{q})e^{\pm i(q^+R_z+\senk{q}\cdot\senk{R})/2}
                                       +a^\dagger_k(\ub{q})e^{\mp i(q^+R_z+\senk{q}\cdot\senk{R})/2}\right]|G^\pm_j\rangle.
\nonumber
\eea
Details of the reduction for the $\Pminus_{\rm n.p.}$ projection can be found
in Appendix~\ref{sec:AppendixA}.  With the combination of all of these terms, the projected
single-source eigenvalue problem becomes
\bea  \label{eq:eigenvalueproblem}
\lefteqn{C^\pm_i\left(\frac{m_i^2}{2m}+\frac{m}{2}\right)|G^\pm_i\rangle
+\frac12 C^\pm_i  \sum_k  \int d\ub{q} \left[\frac{\mu_k^2+q_\perp^2}{q^+}+q^+\right] 
 a_k^\dagger(\ub{q})G^\pm_{ki}(\ub{q})|G^\pm_i\rangle}&& \\
&&+\frac12 g_0 \beta_i \sum_{jk} s_j \beta_j\xi_k C^\pm_j \frac{m_i+m_j}{m}
\int \frac{d\ub{q}}{\sqrt{16\pi^3 q^+}}\left[r_k G^\pm_{kj}(\ub{q})e^{\pm i(q^+R_z+\senk{q}\cdot\senk{R})/2}\right. \nonumber \\
&& \rule{3in}{0mm}                                       
+\left.a^\dagger_k(\ub{q})e^{\mp i(q^+R_z+\senk{q}\cdot\senk{R})/2}\right]|G^\pm_j\rangle \nonumber \\
&& =E^\pm C^\pm_i|G^\pm_i\rangle.  \nonumber
\eea

For this to hold, the coefficient of $a^\dagger_k(\ub{q})$ must be zero,
to remove states with additional particles from the left-hand side:
\bea
0&=&\frac12 C^\pm_i  \left[\frac{\mu_k^2+q_\perp^2}{q^+}+q^+\right] 
 G^\pm_{ki}(\ub{q})|G^\pm_i\rangle \\
&&+\frac12 g_0 \beta_i \sum_j s_j \beta_j\xi_k C^\pm_j \frac{m_i+m_j}{m}
\frac{1}{\sqrt{16\pi^3 q^+}}                  
e^{\mp i(q^+R_z+\senk{q}\cdot\senk{R})/2}|G^\pm_j\rangle. \nonumber
\eea
A slight rearrangement yields an implicit expression for $G^\pm_{ki}$
\be \label{eq:Gki}
C^\pm_i G^\pm_{ki}(\ub{q})|G^\pm_i\rangle 
= -\frac{g_0 \beta_i \xi_k}{\sqrt{16\pi^3 q^+}} 
\frac{e^{\mp i(q^+R_z+\senk{q}\cdot\senk{R})/2}}{\frac{\mu_k^2+q_\perp^2}{q^+}+q^+}
\sum_j s_j \beta_j C^\pm_j \frac{m_i+m_j}{m}|G^\pm_j\rangle.
\ee
The eigenvalue problem (\ref{eq:eigenvalueproblem}) reduces to
\bea 
\lefteqn{C^\pm_i\left(\frac{m_i^2}{2m}+\frac{m}{2}\right)|G^\pm_i\rangle}&& \\
&&+\frac12 g_0 \beta_i \sum_{jk} s_j \beta_j\xi_k C^\pm_j \frac{m_i+m_j}{m}
\int \frac{d\ub{q}}{\sqrt{16\pi^3 q^+}}r_k G^\pm_{kj}(\ub{q})e^{\pm i(q^+R_z+\senk{q}\cdot\senk{R})/2}|G^\pm_j\rangle
 =E^\pm C^\pm_i|G^\pm_i\rangle. \nonumber
\eea
On substitution of the expression for $G^\pm_{ki}$ this becomes
\be \label{eq:EVP1}
C^\pm_i\left(\frac{m_i^2}{2m}+\frac{m}{2}\right)|G^\pm_i\rangle
-\frac{g_0^2}{2} \mu I \beta_i \sum_j s_j \beta_j^2 \frac{m_i+m_j}{m}
\sum_{j'} s_{j'} \beta_{j'} C^\pm_{j'} \frac{m_j+m_{j'}}{m}|G^\pm_{j'}\rangle
 =E^\pm C^\pm_i|G^\pm_i\rangle,
\ee
with $I$ the dimensionless self-energy integral
\be \label{eq:I}
I\equiv \int \frac{d\ub{q}}{16\pi^3\mu} \sum_k \frac{r_k \xi_k^2}{(q^+)^2+q_\perp^2+\mu_k^2}
\ee
and $\mu$ the physical mass of the boson.
The constraint $\sum_k r_k\xi_k^2=0$ on the $\xi_k$ factors makes $I$ finite.

This defines an $n_f\times n_f$ matrix problem
\be \label{eq:FockSpaceEqn}
\left(\frac{m_0^2}{2m}+\frac{m}{2}\right)C^\pm_i|G^\pm_i\rangle
+\sum_j V_{ij}C^\pm_j|G^\pm_j\rangle=E^\pm C^\pm_i|G^\pm_i\rangle,
\ee
where $n_f$ is the number of fermion types and
\be
V_{ij}=-\frac{g_0^2}{2} \mu I \beta_i s_j \beta_j 
\sum_{j'} s_{j'} \beta_{j'}^2 \frac{m_i+m_{j'}}{m}  \frac{m_{j'}+m_j}{m}.
\ee
Now we convert the Fock-space equation (\ref{eq:FockSpaceEqn}) into an algebraic
equation by projecting it onto $\langle G_i^\pm|$
\be \label{eq:Ci}
\left(\frac{m_0^2}{2m}+\frac{m}{2}\right)C^\pm_i
+\sum_j V_{ij}\zeta^\pm_{ij} C^\pm_j=E^\pm C^\pm_i,
\ee
given the overlap integrals
\be \label{eq:zetai}
\zeta^\pm_{ij}=\langle G^\pm_i|G^\pm_j\rangle=\sqrt{Z_iZ_j}
\exp\left(-\sum_k r_k\int d\ub{q}\, G^{\pm*}_{ki}(\ub{q})G^\pm_{kj}(\ub{q})\right),
\ee
with $\zeta^\pm_{ii}=1$ and $\zeta^{\pm*}_{ij}=\zeta^\pm_{ji}$. 
The nontrivial $\zeta^\pm_{ij}$ can be computed from nonlinear
equations for self consistency with the solution for $G^\pm_{ki}$,
which arises in the projection of (\ref{eq:Gki}) onto $\langle G^\pm_i|$:
\be \label{eq:CiGki}
C^\pm_i G^\pm_{ki}(\ub{q})
= -\frac{g_0 \beta_i \xi_k}{\sqrt{16\pi^3 q^+}} 
\frac{e^{\mp i(q^+R_z+\senk{q}\cdot\senk{R})/2}}{\frac{\mu_k^2+q_\perp^2}{q^+}+q^+}
\sum_j s_j \beta_j C^\pm_j \frac{m_i+m_j}{m}\zeta^\pm_{ij}.
\ee
The two equations (\ref{eq:Ci}) and (\ref{eq:zetai}) must then be solved
simultaneously, with $G_{ki}$ in (\ref{eq:zetai}) given by (\ref{eq:CiGki}).
The only $\vec{R}$ dependence appears in the exponentials which 
cancel in (\ref{eq:zetai}), leaving the equations independent of $\vec{R}$.
Therefore, the ground state determines the physical fermion
mass $m=E^\pm$ independent of the source location $\pm\vec{R}/2$,
and the eigenvalue problem (\ref{eq:Ci}) provides the renormalization of the bare
mass $m_0$ implicitly by giving $m$ as a function of $m_0$.  
We then use these solutions to construct a solution
for the two-source case in the next section.  For this purpose,
an explicit solution of the (\ref{eq:Ci})-(\ref{eq:zetai}) system
will not be needed.

\section{Two static sources}  \label{sec:Double}

To compute the effective potential between two sources
a distance $R$ apart, we place them at $\ub{x}=(\mp R_z,\pm\senk{R}/2)$
and construct the eigenstate of the ordinary energy ${\cal E}$.
The effective potential is then the difference between the
eigenvalue and the total rest mass $2m$, with $m$ specified 
by the single-source problem solved in the previous section.
Following the case of scalar Yukawa theory~\cite{ScalarYukawa},
we construct an {\em ansatz} for the eigenstate as a product
of single-source solutions:
\be
|G^+G^-F^+F^-;s_1s_2\rangle\equiv\sum_{ij}C_{ij}|G^+_iF^+_{is_1}\rangle|G^-_jF^-_{js_2}\rangle.
\ee
This is to be a solution of
\be
{\cal E}|G^+G^-F^+F^-;s_1s_2\rangle=E|G^+G^-F^+F^-;s_1s_2\rangle.
\ee

We proceed as before with a projection onto the static fermion
states
\be  \label{eq:twosource}
s_i s_j\langle F^-_{js_2}|\langle F^+_{is_1}|{\cal E}|G^+G^-F^+F^-;s_1s_2\rangle
=Es_i s_j\langle F^-_{js_2}|\langle F^+_{is_1}|G^+G^-F^+F^-;s_1s_2\rangle.
\ee
Cross terms between $F^+$ and $F^-$ do not contribute because they
are proportional to 
\be
\int d\ub{p} F^{+*}(\ub{p})F^-(\ub{p})=\int \frac{d\ub{x}}{16\pi^3}\psi^{+*}(\ub{x})\psi^-(\ub{x}),
\ee
and the second integral is zero from the lack of overlap between
narrow wave packets centered apart.  Again, the spin-flip terms of 
$\Pminus_{\rm n.p.}$ do not contribute, being proportional
to the transverse fermion momentum $\senk{p}$ for which the expectation value is zero.
The right-hand side of (\ref{eq:twosource}) is
\be
Es_i s_j\langle F^-_{js_2}|\langle F^+_{is_1}|G^+G^-F^+F^-;s_1s_2\rangle=EC_{ij}|G^+_i\rangle|G^-_j\rangle.
\ee
The projected terms in ${\cal E}$ for the left-hand side are
\bea
\lefteqn{s_i s_j\langle F^-_{js_2}|\langle F^+_{is_1}|\frac12(\Pminus_{0f}+\Pplus_f)|G^+G^-F^+F^-;s_1s_2\rangle}&& \\
&&=\frac12 C_{ij}\int d\ub{p} \left\{\left[\frac{m_i^2+p_\perp^2}{p^+}+p^+\right] |F_i^+(\ub{p})|^2
                                     +\left[\frac{m_j^2+p_\perp^2}{p^+}+p^+\right] |F_j^+(\ub{p})|^2\right\}|G^+_i\rangle|G^-_j\rangle 
                                     \nonumber \\
&&= C_{ij}\left(\frac{m_i^2}{2m}+\frac{m_j^2}{2m}+m\right)|G^+_i\rangle|G^-_j\rangle, \nonumber
\eea
\bea \label{eq:projectedP0a}
\lefteqn{s_i s_j\langle F^-_{js_2}|\langle F^+_{is_1}|\frac12(\Pminus_{0a}+\Pplus_a)|G^+G^-F^+F^-;s_1s_2\rangle}&& \\
&&=\frac12 C_{ij}  \sum_k  \int d\ub{q} \left[\frac{\mu_k^2+q_\perp^2}{q^+}+q^+\right] 
 a_k^\dagger(\ub{q})a_k(\ub{q})|G^+_i\rangle|G^-_j\rangle, \nonumber  
\eea
and
\bea \label{eq:projectedPnp2}
\lefteqn{s_i s_j\langle F^-_{js_2}|\langle F^+_{is_1}|\frac12\Pminus_{\rm n.p.}|G^+G^-F^+F^-;s_1s_2\rangle
=  \frac12 g_0\sum_k\xi_k \int\frac{d\ub{q}}{\sqrt{16\pi^3 q^+}}}&& \\
&&  \times  \left\{
\beta_i \sum_{i'} s_{i'}\beta_{i'}  C_{i'j}
   \left[a_k(\ub{q})e^{i(q^+R_z+\senk{q}\cdot\senk{R})/2}+a^\dagger_k(\ub{q})e^{-i(q^+R_z+\senk{q}\cdot\senk{R})/2}\right]
   \frac{m_i+m_{i'}}{m}|G^+_{i'}\rangle|G^-_j\rangle  \right.
\nonumber \\
&&  \left. + \beta_j \sum_{j'} s_{j'}\beta_{j'} C_{ij'}
   \left[a_k(\ub{q})e^{-i(q^+R_z+\senk{q}\cdot\senk{R})/2}+a^\dagger_k(\ub{q})e^{i(q^+R_z+\senk{q}\cdot\senk{R})/2}\right]
  \frac{m_j+m_{j'}}{m}|G^+_i\rangle|G^-_{j'}\rangle    \right\}.  \nonumber
\eea
Details of this last projection are again in Appendix~\ref{sec:AppendixA}.  The boson annihilation operators 
in (\ref{eq:projectedP0a}) and (\ref{eq:projectedPnp2}) can be replaced with use
of $a_k(\ub{q})|G^+_i\rangle|G^-_j\rangle=r_k[G^+_{ki}(\ub{q})+G^-_{kj}(\ub{q})]|G^+_i\rangle|G^-_j\rangle$,
which is the two-source extension of (\ref{eq:aG}), where a coherent state is an
eigenstate of the annihilation operator.

The projected eigenvalue problem is
\bea  \label{eq:twosourceEVP}
\lefteqn{C_{ij}\left\{\frac{m_i^2}{2m}+\frac{m_j^2}{2m}+m
+\frac12 \sum_k  \int d\ub{q} \left[\frac{\mu_k^2+q_\perp^2}{q^+}+q^+\right] 
 a_k^\dagger(\ub{q})r_k[G^+_{ki}(\ub{q})+G^-_{kj}(\ub{q})]\right\}|G^+_i\rangle|G^-_j\rangle}&& \nonumber \\
&& + \frac12 g_0\sum_k\xi_k \int\frac{d\ub{q}}{\sqrt{16\pi^3 q^+}}
\left\{
\beta_i \sum_{i'} s_{i'}\beta_{i'}  C_{i'j}
   \left[r_k[G^+_{ki'}(\ub{q})+G^-_{kj}(\ub{q})]e^{i(q^+R_z+\senk{q}\cdot\senk{R})/2} \right. \right.  \\
&& \rule{2in}{0mm}\left. +a^\dagger_k(\ub{q})e^{-i(q^+R_z+\senk{q}\cdot\senk{R})/2}\right]
    \frac{m_i+m_{i'}}{m}|G^+_{i'}\rangle|G^-_j\rangle \nonumber \\
&&  + \beta_j \sum_{j'} s_{j'}\beta_{j'} C_{ij'}
   \left[r_k[G^+_{ki}(\ub{q})+G^-_{kj'}(\ub{q})]e^{-i(q^+R_z+\senk{q}\cdot\senk{R})/2}\right. \nonumber  \\
&& \rule{1in}{0mm} \left.\left. +a^\dagger_k(\ub{q})e^{i(q^+R_z+\senk{q}\cdot\senk{R})/2}\right]
    \frac{m_j+m_{j'}}{m}|G^+_i\rangle|G^-_{j'}\rangle    \right\} 
=E C_{ij}|G^+_i\rangle|G^-_j\rangle.
\nonumber
\eea
The reduction of this eigenvalue problem is detailed in Appendix~\ref{sec:AppendixB}.

From the reduction we obtain the effective potential $V_{\rm eff}\equiv E-2m$ as
\be \label{eq:VeffFinal}
V_{\rm eff}=-\frac{g_0^2}{2}  \left(\sum_i \frac{s_i\beta_i^2 m_i}{m}\right)^2 \frac{e^{-\mu R}}{8 \pi R},
\ee
which is clearly rotationally invariant.
We define the physical coupling $g$ by a match to the standard form 
for the Yukawa potential $-\frac{g^2}{4\pi R}e^{-\mu R}$, which implies
\be
g=g_0\sum_i \frac{s_i\beta_i^2 m_i}{2m}.
\ee
We can, of course, have $g=g_0$ if we include 2 PV fermions and
impose the additional constraint $\sum_i s_i\beta_i^2m_i=2m$.

\section{Summary}  \label{sec:Summary}

In this work we have thus obtained the standard, rotationally invariant
Yukawa potential as the effective potential between two static sources
in quenched, light-front Yukawa theory.  The effective potential comes
from the interference between the two boson clouds that dress the
individual sources and is computed nonperturbatively.
The rotational invariance exists
despite the special status for the $z$ axis in light-front quantization.

The key to our approach is to recognize the ordinary energy as
the relevant quantity, both because momentum is not conserved
when sources are static and because the effective potential
should be defined in terms of this energy.  Light-front energy
combines energy and a momentum component, making it only indirectly
related.

To carry out the calculation, we have introduced Pauli-Villars
fermions and bosons.  The PV fermions eliminate instantaneous
interaction terms which would otherwise interfere with the
construction of analytic solutions.  The PV bosons regulate the
infinite self-energy of the sources.  The couplings of these
are adjusted to satisfy constraints that guarantee the regularization,
the correct mass and coupling renormalizations, and the removal
of instantaneous fermion interactions from the light-front 
Hamiltonian.  The instantaneous interactions are restored in
the limit of infinite PV mass.

Given the successful derivation of a rotationally invariant
potential in quenched Yukawa theory, the
next step to be taken is to introduce
pair production and annihilation of free fermions and their
own accompanying PV counterparts.  (The PV fermions associated
with the static sources need to be separate because they
are themselves static.)  With pairs included in the basis,
a coherent state solution will no longer be available as 
the full solution, and Fock-space methods must be invoked.
To have an eigenvalue problem of finite size will then
require truncation, either explicitly in Fock space or
in the operator sense of the light-front coupled-cluster
method~\cite{LFCC}.  The effects of pairs will include
renormalization of the boson mass, renormalization of
the static fermion coupling, and modifications of the
form of the effective potential.  At short separations,
these modifications will be due to the charge 
renormalization and the screening that takes place.
At large separations, pairs provide for the Yukawa
analog of string breaking.  Completion of this work
in Yukawa theory will provide a useful reference
for the calculation of effective potentials in QED and QCD.


\appendix

\section{Projections for the no-pair contribution}  \label{sec:AppendixA}

We first consider the single-source case.
Substitution of the definitions of the individual factors
$|F^\pm_{is}\rangle$, $\Pminus_{\rm n.p.}$, and $|G^\pm F^\pm; s\rangle$,
as given in (\ref{eq:Fpm}), (\ref{eq:Pnp}), and (\ref{eq:GFs}), respectively,
yields
\bea
\lefteqn{s_i\langle F^\pm_{is}|\frac12\Pminus_{\rm n.p.}|G^\pm F^\pm; s\rangle
=\frac12 g_0 s_i \int d\ub{p}' F^{\pm*}(\ub{p}') _f\langle 0|b_{is}(\ub{p}')
\sum_{i'jk} \beta_{i'} \beta_j\xi_k\int d\ub{x}}&&\\
&& \times \int \frac{d\ub{q}}{\sqrt{16\pi^3 q^+}}\left[a_k(\ub{q})e^{-i\ub{q}\cdot\ub{x}}+a^\dagger(\ub{q})e^{i\ub{q}\cdot\ub{x}}\right]
\int \frac{d\ub{p}_1 d\ub{p}_2}{16 \pi^3}\left(\frac{m_{i'}}{p_1^+}+\frac{m_j}{p_2^+}\right)
e^{i(\ub{p}_1-\ub{p}_2)\cdot\ub{x}}\sum_{s'}b^\dagger_{i's'}(\ub{p}_1)b_{js'}(\ub{p}_2) \nonumber \\
&&\times\sum_{i''}C^\pm_{i''}\int d\ub{p} F^\pm(\ub{p})b^\dagger_{i''s}(\ub{p})\vacf|G^\pm_{i''}\rangle.
\nonumber
\eea
Given the contractions 
\be
b_{is}(\ub{p}')b^\dagger_{i's'}(\ub{p}_1)\rightarrow s_i\delta_{ii'}\delta_{ss'}\delta(\ub{p}'-\ub{p}_1)
\ee
and 
\be
b_{js'}(\ub{p}_2)b^\dagger_{i''s}(\ub{p})\rightarrow s_j\delta_{ji''}\delta_{s's}\delta(\ub{p}-\ub{p}_2),
\ee
the Kronecker and Dirac deltas and the property $s_i^2=1$ can be used to reduce the expression to
\bea
\lefteqn{s_i\langle F^\pm_{is}|\frac12\Pminus_{\rm n.p.}|G^\pm F^\pm; s\rangle
=\frac12 g_0 \beta_i \sum_{jk} s_j \beta_j\xi_k
\int d\ub{x} \int \frac{d\ub{p}'}{\sqrt{16\pi^3}} F^{\pm*}(\ub{p}') e^{i\ub{p}'\cdot\ub{x}}}&& \\
&& \times \int \frac{d\ub{q}}{\sqrt{16\pi^3 q^+}}\left[a_k(\ub{q})e^{-i\ub{q}\cdot\ub{x}}+a^\dagger(\ub{q})e^{i\ub{q}\cdot\ub{x}}\right]
\int\frac{d\ub{p}}{\sqrt{16\pi^3}} F^\pm(\ub{p})e^{-i\ub{p}\cdot\ub{x}}
\left(\frac{m_i}{p^{\prime +}}+\frac{m_j}{p^+}\right) C^\pm_j |G^\pm_j\rangle.
\nonumber
\eea
The momentum-space wave function $F^\pm$ is peaked at $(m,\senk{0})$,
which allows any factor of $p^+$ or $p^{\prime +}$ to be replaced with $m$.
The two Fourier transforms of $F^\pm$
can be written in terms of the spatial wave function $\psi$ in (\ref{eq:psi})
and the product replaced by the coordinate-space delta functions of (\ref{eq:source}):
\be
\int \frac{d\ub{p}'}{\sqrt{16\pi^3}} F^{\pm*}(\ub{p}') e^{i\ub{p}'\cdot\ub{x}}
\int\frac{d\ub{p}}{\sqrt{16\pi^3}} F^\pm(\ub{p})e^{-i\ub{p}\cdot\ub{x}}=|\psi^\pm(\ub{x})|^2
\rightarrow \delta(x^-\pm R_z)\delta(\senk{x}\mp\senk{R}/2).
\ee
Integration over $\ub{x}$ then gives
\bea
\lefteqn{s_i\langle F^\pm_{is}|\frac12\Pminus_{\rm n.p.}|G^\pm F^\pm; s\rangle
=\frac12 g_0 \beta_i \sum_{jk} s_j \beta_j\xi_k C^\pm_j\frac{m_i+m_j}{m}}&& \\
&&\times \int \frac{d\ub{q}}{\sqrt{16\pi^3 q^+}}\left[a_k(\ub{q})e^{\pm i(q^+R_z+\senk{q}\cdot\senk{R})/2}
        +a^\dagger(\ub{q})e^{\mp i(q^+R_z+\senk{q}\cdot\senk{R})/2}\right]
|G^\pm_j\rangle.
\nonumber
\eea
Use of (\ref{eq:aG}) then yields the result given in (\ref{eq:projectedPnp}).

When there are two sources, the projection needed is
\bea
\lefteqn{s_i s_j\langle F^-_{js_2}|\langle F^+_{is_1}|\frac12\Pminus_{\rm n.p.}|G^+G^-F^+F^-;s_1s_2\rangle}&& \\
&&= \frac12 s_i s_j \, \mbox{}_f\langle0|\int d\ub{p}'_2 F^{-*}(\ub{p}'_2)b_{js_2}(\ub{p}'_2)
\int d\ub{p}'_1 F^{+*}(\ub{p}'_1)b_{is_1}(\ub{p}'_1)
g_0\sum_{lmk}\beta_l\beta_m\xi_k\int d\ub{x} \nonumber  \\
&& \times \int\frac{d\ub{q}}{\sqrt{16\pi^3 q^+}}
   \left[a_k(\ub{q})e^{-i\ub{q}\cdot\ub{x}}+a^\dagger_k(\ub{q})e^{i\ub{q}\cdot\ub{x}}\right]
   \int \frac{d\ub{p}_1 d\ub{p}_2}{16 \pi^3}e^{i(\ub{p}_1-\ub{p}_2)\cdot\ub{x}}\left(\frac{m_l}{p^+_1}+\frac{m_m}{p^+_2}\right)
   \sum_s b^\dagger_{ls}(\ub{p}_1)b_{ms}(\ub{p}_2) \nonumber \\
&&  \times \sum_{i'j'} C_{i'j'}
   \int d\ub{p}''_1 F^+(\ub{p}''_1)b^\dagger_{i's_1}(\ub{p}''_1)
\int d\ub{p}''_2 F^-(\ub{p}''_2)b^\dagger_{j's_2}(\ub{p}''_2)\vacf|G^+_{i'}\rangle|G^-_{j'}\rangle.
\nonumber
\eea
Contraction of $b_{ms}(\ub{p}_2)$ with each of the rightmost creation operators yields
\bea
\lefteqn{s_i s_j\langle F^-_{js_2}|\langle F^+_{is_1}|\frac12\Pminus_{\rm n.p.}|G^+G^-F^+F^-;s_1s_2\rangle}&& \\
&&= \frac12 s_i s_j \,\mbox{}_f\langle0|\int d\ub{p}'_2 F^{-*}(\ub{p}'_2)b_{js_2}(\ub{p}'_2)
\int d\ub{p}'_1 F^{+*}(\ub{p}'_1)b_{is_1}(\ub{p}'_1)
g_0\sum_{lmk}\beta_l\beta_m\xi_k\int d\ub{x} \nonumber  \\
&& \times \int\frac{d\ub{q}}{\sqrt{16\pi^3 q^+}}
   \left[a_k(\ub{q})e^{-i\ub{q}\cdot\ub{x}}+a^\dagger_k(\ub{q})e^{i\ub{q}\cdot\ub{x}}\right]
   \int \frac{d\ub{p}_1 d\ub{p}_2}{16 \pi^3}e^{i(\ub{p}_1-\ub{p}_2)\cdot\ub{x}}\left(\frac{m_l}{p^+_1}+\frac{m_m}{p^+_2}\right)
   \sum_s b^\dagger_{ls}(\ub{p}_1) \nonumber \\
&&  \times \sum_{i'j'} C_{i'j'} \left\{
   \int d\ub{p}''_1 F^+(\ub{p}''_1)
\int d\ub{p}''_2 F^-(\ub{p}''_2)
  s_m\delta_{ss_1}\delta_{mi'}\delta(\ub{p}_2-\ub{p}''_1)b^\dagger_{j's_2}(\ub{p}''_2)\right.
\nonumber \\
&&  \left. -\int d\ub{p}''_1 F^+(\ub{p}''_1)b^\dagger_{i's_1}(\ub{p}''_1)
\int d\ub{p}''_2 F^-(\ub{p}''_2)
s_m\delta_{ss_2}\delta_{mj'}\delta(\ub{p}_2-\ub{p}''_2)\right\}\vacf|G^+_{i'}\rangle|G^-_{j'}\rangle.
\nonumber
\eea
With use of the Kronecker and Dirac deltas, this becomes
\bea
\lefteqn{s_i s_j\langle F^-_{js_2}|\langle F^+_{is_1}|\frac12\Pminus_{\rm n.p.}|G^+G^-F^+F^-;s_1s_2\rangle}&& \\
&&= \frac12 s_i s_j \,\mbox{}_f\langle0|\int d\ub{p}'_2 F^{-*}(\ub{p}'_2)b_{js_2}(\ub{p}'_2)
\int d\ub{p}'_1 F^{+*}(\ub{p}'_1)b_{is_1}(\ub{p}'_1)
g_0\sum_{lk}\beta_l\xi_k\int d\ub{x} \nonumber  \\
&& \times \int\frac{d\ub{q}}{\sqrt{16\pi^3 q^+}}
   \left[a_k(\ub{q})e^{-i\ub{q}\cdot\ub{x}}+a^\dagger_k(\ub{q})e^{i\ub{q}\cdot\ub{x}}\right]
   \int \frac{d\ub{p}_1 d\ub{p}_2}{16 \pi^3}e^{i(\ub{p}_1-\ub{p}_2)\cdot\ub{x}}
  \nonumber \\
&&  \times \sum_{i'j'} C_{i'j'} \left\{s_{i'}\beta_{i'}\left(\frac{m_l}{p^+_1}+\frac{m_{i'}}{p^+_2}\right)
   F^+(\ub{p}_2)
\int d\ub{p}''_2 F^-(\ub{p}''_2)
   b^\dagger_{ls_1}(\ub{p}_1)b^\dagger_{j's_2}(\ub{p}''_2)\right.
\nonumber \\
&&  \left. -s_{j'}\beta_{j'}\left(\frac{m_l}{p^+_1}+\frac{m_{j'}}{p^+_2}\right)
\int d\ub{p}''_1F^+(\ub{p}''_1)b^\dagger_{ls_2}(\ub{p}_1)b^\dagger_{i's_1}(\ub{p}''_1)
F^-(\ub{p}_2)\right\}\vacf|G^+_{i'}\rangle|G^-_{j'}\rangle.
\nonumber
\eea
The remaining contractions produce
\bea
\lefteqn{s_i s_j\langle F^-_{js_2}|\langle F^+_{is_1}|\frac12\Pminus_{\rm n.p.}|G^+G^-F^+F^-;s_1s_2\rangle}&& \\
&&= \frac12 s_i s_j \int d\ub{p}'_2 F^{-*}(\ub{p}'_2)
\int d\ub{p}'_1 F^{+*}(\ub{p}'_1)
g_0\sum_{lk}\beta_l\xi_k\int d\ub{x} \nonumber  \\
&& \times \int\frac{d\ub{q}}{\sqrt{16\pi^3 q^+}}
   \left[a_k(\ub{q})e^{-i\ub{q}\cdot\ub{x}}+a^\dagger_k(\ub{q})e^{i\ub{q}\cdot\ub{x}}\right]
   \int \frac{d\ub{p}_1 d\ub{p}_2}{16 \pi^3}e^{i(\ub{p}_1-\ub{p}_2)\cdot\ub{x}}
  \nonumber \\
&&  \times \sum_{i'j'} C_{i'j'} \left\{s_{i'}\beta_{i'}\left(\frac{m_l}{p^+_1}+\frac{m_{i'}}{p^+_2}\right)
   F^+(\ub{p}_2)
\int d\ub{p}''_2 F^-(\ub{p}''_2)\right. \nonumber \\
&& \rule{0.25in}{0mm} \times
\left[s_i \delta_{il}\delta(\ub{p}'_1-\ub{p}_1) s_j \delta_{jj'}\delta(\ub{p}'_2-\ub{p}''_2)
     -s_i \delta_{s_1s_2}\delta_{ij'}\delta(\ub{p}'_1-\ub{p}''_2) s_j \delta_{s_2s_1}\delta_{jl}\delta(\ub{p}'_2-\ub{p}_1)\right]
\nonumber \\
&&  \rule{0.1in}{0mm} -s_{j'}\beta_{j'}\left(\frac{m_l}{p^+_1}+\frac{m_{j'}}{p^+_2}\right)
\int d\ub{p}''_1 F^+(\ub{p}''_1) F^-(\ub{p}_2) \nonumber \\
&& \rule{0.25in}{0mm} \times \left.
\left[s_i \delta_{s_1s_2}\delta_{il}\delta(\ub{p}'_1-\ub{p}_1) s_j\delta_{s_2s_1}\delta_{ji'}\delta(\ub{p}'_2-\ub{p}''_1)
       -s_i \delta_{ii'}\delta(\ub{p}'_1-\ub{p}''_1) s_j \delta_{jl}\delta(\ub{p}'_2-\ub{p}_1)\right] 
       \right\}|G^+_{i'}\rangle|G^-_{j'}\rangle.
\nonumber
\eea
The additional Kronecker and Dirac deltas, and the fact that $s_i^2=1$, reduce this to
\bea
\lefteqn{s_i s_j\langle F^-_{js_2}|\langle F^+_{is_1}|\frac12\Pminus_{\rm n.p.}|G^+G^-F^+F^-;s_1s_2\rangle}&& \\
&&=  \frac12 g_0\sum_k\xi_k\int d\ub{x}
\int\frac{d\ub{q}}{\sqrt{16\pi^3 q^+}}
   \left[a_k(\ub{q})e^{-i\ub{q}\cdot\ub{x}}+a^\dagger_k(\ub{q})e^{i\ub{q}\cdot\ub{x}}\right]
   \int \frac{d\ub{p}_1 d\ub{p}_2}{16 \pi^3}e^{i(\ub{p}_1-\ub{p}_2)\cdot\ub{x}} \nonumber \\
&&  \times  \left\{
\sum_{i'} s_{i'}\beta_{i'}
\left[ C_{i'j}\beta_i 
\int d\ub{p}'_2\,|F^-(\ub{p}'_2)|^2 F^{+*}(\ub{p}_1) F^+(\ub{p}_2)
\left(\frac{m_i}{p^+_1}+\frac{m_{i'}}{p^+_2}\right)
   |G^+_{i'}\rangle|G^-_j\rangle \right. \right. \\
&& \left.- \delta_{s_1s_2} C_{i'i}\beta_j 
\int d\ub{p}'_1\,F^{+*}(\ub{p}'_1)F^-(\ub{p}'_1) F^{-*}(\ub{p}_1) F^+(\ub{p}_2)
\left(\frac{m_j}{p^+_1}+\frac{m_{i'}}{p^+_2}\right)
|G^+_{i'}\rangle|G^-_i\rangle \right]
\nonumber \\
&&  -\sum_{j'} s_{j'}\beta_{j'}
 \left[ \delta_{s_1s_2} C_{jj'} \beta_i 
\int d\ub{p}'_2\,F^{-*}(\ub{p}'_2)F^+(\ub{p}'_2)F^{+*}(\ub{p}_1) F^-(\ub{p}_2)
\left(\frac{m_i}{p^+_1}+\frac{m_{j'}}{p^+_2}\right)
 |G^+_j\rangle|G^-_{j'}\rangle \right.
\nonumber \\
&& \left. \left.      - C_{ij'} \beta_j 
\int d\ub{p}'_1\,|F^+(\ub{p}'_1)|^2 F^{-*}(\ub{p}_1) F^-(\ub{p}_2)
\left(\frac{m_j}{p^+_1}+\frac{m_{j'}}{p^+_2}\right)
|G^+_i\rangle|G^-_{j'}\rangle \right]   \right\}. \nonumber
\eea
The $\ub{p}'_1$ and $\ub{p}'_2$ integrations yield
either unity for direct terms or zero for cross terms;
the latter can be identified by the leading $\delta_{s_1s_2}$.
The integral in the direct terms is the normalization
integral for $F^\pm$.  The integral for the cross
terms is zero because of zero overlap between the
wave packets of the two sources.  The integrals over
$\ub{p}_1$ and $\ub{p}_2$ can be rewritten in terms
of the Fourier transforms $\psi^\pm$, with factors
of $\ub{p}'_1$ and $\ub{p}'_2$ replaced with $(m,\senk{0})$
at the peak of $F^\pm$.  This leaves factors of 
$|\psi^\pm(\ub{x})|^2$ which become
$\delta(x^-\pm R_z)\delta(\senk{x}\mp\senk{R}/2)$.
The integral over $\ub{x}$ can then be performed.
The resulting expression is what is quoted in
(\ref{eq:projectedPnp2}).

\section{Reduction of the two-source eigenvalue problem}  \label{sec:AppendixB}

We begin from the statement of the full eigenvalue problem for
two static sources as given in (\ref{eq:twosourceEVP}).
The coefficient of the collected $a^\dagger_k(\ub{q})$ terms is
\bea
\lefteqn{C_{ij}\frac12 \sum_k   \left[\frac{\mu_k^2+q_\perp^2}{q^+}+q^+\right] 
 r_k[G^+_{ki}(\ub{q})+G^-_{kj}(\ub{q})]|G^+_i\rangle|G^-_j\rangle}&& \\
&& + \frac12 g_0\xi_k \frac{1}{\sqrt{16\pi^3 q^+}}
\left\{
\beta_i \sum_{i'} s_{i'}\beta_{i'}  C_{i'j}
   e^{-i(q^+R_z+\senk{q}\cdot\senk{R})/2}
   \frac{m_i+m_{i'}}{m}|G^+_{i'}\rangle|G^-_j\rangle \right.
\nonumber \\
&& \rule{0.5in}{0mm} \left. + \beta_j \sum_{j'} s_{j'}\beta_{j'} C_{ij'}
   e^{i(q^+R_z+\senk{q}\cdot\senk{R})/2}
    \frac{m_j+m_{j'}}{m}|G^+_i\rangle|G^-_{j'}\rangle  \right\}.
\nonumber
\eea
If we set $C_{ij}=C^+_iC^-_j$, this coefficient of $a^\dagger_k(\ub{q})$ is automatically zero,
given the solution to the single-source case. This leaves the
double-source eigenvalue problem  (\ref{eq:twosourceEVP}) in the form
\bea
\lefteqn{C^+_iC^-_j\left\{\frac{m_i^2}{2m}+\frac{m_j^2}{2m}+m\right\}|G^+_i\rangle|G^-_j\rangle
+ \frac12 g_0\sum_k\xi_k \int\frac{d\ub{q}}{\sqrt{16\pi^3 q^+}}}&& \\
&& \times \left\{
\beta_i \sum_{i'} s_{i'}\beta_{i'}  C^+_{i'}C^-_j
   r_k[G^+_{ki'}(\ub{q})+G^-_{kj}(\ub{q})]e^{i(q^+R_z+\senk{q}\cdot\senk{R})/2}
 \frac{m_i+m_{i'}}{m}|G^+_{i'}\rangle|G^-_j\rangle \right.
\nonumber \\
&& \rule{0.5in}{0mm} \left. + \beta_j \sum_{j'} s_{j'}\beta_{j'} C^+_{i}C^-_{j'}
   r_k[G^+_{ki}(\ub{q})+G^-_{kj'}(\ub{q})]e^{-i(q^+R_z+\senk{q}\cdot\senk{R})/2}
    \frac{m_j+m_{j'}}{m}|G^+_i\rangle|G^-_{j'}\rangle    \right\}  \nonumber \\
&&=E C^+_iC^-_j|G^+_i\rangle|G^-_j\rangle.
\nonumber
\eea
The combination $C^\pm_i G^\pm_{ki}(\ub{q})|G^\pm_i\rangle$
can be replaced in each appearance with use of (\ref{eq:Gki})
\bea
\lefteqn{C^+_iC^-_j\left\{\frac{m_i^2}{2m}+\frac{m_j^2}{2m}+m\right\}|G^+_i\rangle|G^-_j\rangle
- \frac12 g_0\sum_k\xi_k \int\frac{d\ub{q}}{\sqrt{16\pi^3 q^+}}}&& \\
&& \times \left\{
\beta_i \sum_{i'} s_{i'}\beta_{i'}  C^-_j
   r_k e^{i(q^+R_z+\senk{q}\cdot\senk{R})/2}
   \frac{m_i+m_{i'}}{m} \frac{g_0 \beta_{i'} \xi_k}{\sqrt{16\pi^3 q^+}} 
\frac{e^{- i(q^+R_z+\senk{q}\cdot\senk{R})/2}}{\frac{\mu_k^2+q_\perp^2}{q^+}+q^+} \right. \nonumber \\
&&  \rule{2in}{0mm} \times \sum_{j'} s_{j'} \beta_{j'} C^+_{j'} \frac{m_{i'}+m_{j'}}{m}|G^+_{j'}\rangle
|G^-_j\rangle \nonumber \\
&&+\beta_i \sum_{i'} s_{i'}\beta_{i'}  C^+_{i'}
   r_k  e^{i(q^+R_z+\senk{q}\cdot\senk{R})/2}\frac{m_i+m_{i'}}{m}|G^+_{i'}\rangle 
\frac{g_0 \beta_j \xi_k}{\sqrt{16\pi^3 q^+}} 
\frac{e^{ i(q^+R_z+\senk{q}\cdot\senk{R})/2}}{\frac{\mu_k^2+q_\perp^2}{q^+}+q^+} \nonumber \\
&&  \rule{2in}{0mm} \times
\sum_{j'} s_{j'} \beta_{j'} C^-_{j'} \frac{m_j+m_{j'}}{m}|G^-_{j'}\rangle \nonumber \\
&& + \beta_j \sum_{j'} s_{j'}\beta_{j'} C^-_{j'}
   r_k  e^{-i(q^+R_z+\senk{q}\cdot\senk{R})/2}\frac{m_j+m_{j'}}{m}
\frac{g_0 \beta_i \xi_k}{\sqrt{16\pi^3 q^+}} 
\frac{e^{- i(q^+R_z+\senk{q}\cdot\senk{R})/2}}{\frac{\mu_k^2+q_\perp^2}{q^+}+q^+} \nonumber \\
&&  \rule{2in}{0mm} \times
\sum_{i'} s_{i'} \beta_{i'} C^+_{i'} \frac{m_i+m_{i'}}{m}|G^+_{i'}\rangle|G^-_{j'}\rangle    \nonumber \\
&& + \beta_j \sum_{j'} s_{j'}\beta_{j'} C^+_{i}
   r_k e^{-i(q^+R_z+\senk{q}\cdot\senk{R})/2}\frac{m_j+m_{j'}}{m}|G^+_i\rangle
\frac{g_0 \beta_{j'} \xi_k}{\sqrt{16\pi^3 q^+}} 
\frac{e^{ i(q^+R_z+\senk{q}\cdot\senk{R})/2}}{\frac{\mu_k^2+q_\perp^2}{q^+}+q^+}  \nonumber \\
&&  \rule{2in}{0mm} \left. \times
\sum_{i'} s_{i'} \beta_{i'} C^-_{i'} \frac{m_{j'}+m_{i'}}{m}|G^-_{i'}\rangle
\right\}  \nonumber \\
&&=E C^+_iC^-_j|G^+_i\rangle|G^-_j\rangle.
\nonumber
\eea
The $\ub{q}$ dependent factors can be combined into 
a factor that is either the self-energy integral $I$
defined in (\ref{eq:I}), where the exponential factors cancel,
or the integral
\be
Y^\pm(\vec{R})=\int \frac{d\ub{q}}{16\pi^3}\sum_k 
\frac{r_k\xi_k^2 e^{\pm i(q^+R_z+\senk{q}\cdot\senk{R})}}{(q^+)^2+q_\perp^2+\mu_k^2}.
\ee
We then obtain
\bea
\lefteqn{C^+_iC^-_j\left\{\frac{m_i^2}{2m}+\frac{m_j^2}{2m}+m\right\}|G^+_i\rangle|G^-_j\rangle}&& \\
&& - \frac{g_0^2}{2} \left\{
\beta_i C^-_j \mu I \sum_{i'} s_{i'}\beta_{i'}^2 \frac{m_i+m_{i'}}{m}  
 \sum_{j'} s_{j'} \beta_{j'} C^+_{j'} \frac{m_{i'}+m_{j'}}{m}|G^+_{j'}\rangle
|G^-_j\rangle \right. \nonumber \\
&&+\beta_i \beta_j Y^+(\vec{R}) \sum_{i'} s_{i'}\beta_{i'}  C^+_{i'} \frac{m_i+m_{i'}}{m} 
\sum_{j'} s_{j'} \beta_{j'} C^-_{j'} \frac{m_j+m_{j'}}{m} |G^+_{i'}\rangle |G^-_{j'}\rangle \nonumber \\
&& + \beta_i \beta_j  Y^-(\vec{R})  \sum_{j'} s_{j'}\beta_{j'} C^-_{j'}\frac{m_j+m_{j'}}{m}
\sum_{i'} s_{i'} \beta_{i'} C^+_{i'} \frac{m_i+m_{i'}}{m}|G^+_{i'}\rangle|G^-_{j'}\rangle    \nonumber \\
&& \left. +  C^+_i \beta_j \mu I \sum_{j'} s_{j'}\beta_{j'}^2\frac{m_j+m_{j'}}{m}
\sum_{i'} s_{i'} \beta_{i'} C^-_{i'} \frac{m_{j'}+m_{i'}}{m}|G^+_i\rangle|G^-_{i'}\rangle
\right\}  \nonumber \\
&&=E C^+_iC^-_j|G^+_i\rangle|G^-_j\rangle.
\nonumber
\eea
This can be rearranged to reveal parts directly related to the 
single-source problem
\bea
\lefteqn{\left\{C^+_i\left(\frac{m_i^2}{2m}+\frac{m}{2}\right)|G^+_i\rangle
- \frac{g_0^2}{2}
\beta_i  \mu I \sum_{i'} s_{i'}\beta_{i'}^2 \frac{m_i+m_{i'}}{m}   
 \sum_{j'} s_{j'} \beta_{j'} C^+_{j'} \frac{m_{i'}+m_{j'}}{m}|G^+_{j'}\rangle\right\}
C^-_j|G^-_j\rangle}&& \nonumber \\
&&+C^+_i|G^+_i\rangle\left\{C^-_j\left(\frac{m_j^2}{2m}+\frac{m}{2}\right)|G^-_j\rangle
- \frac{g_0^2}{2}
\beta_j \mu I \sum_{j'} s_{j'}\beta_{j'}^2 \frac{m_j+m_{j'}}{m}
\sum_{i'} s_{i'} \beta_{i'} C^-_{i'} \frac{m_{j'}+m_{i'}}{m}|G^-_{i'}\rangle\right\} \nonumber \\
&&- \frac{g_0^2}{2}\beta_i \beta_j Y^+(\vec{R}) \sum_{i'} s_{i'}\beta_{i'}  C^+_{i'} 
\frac{m_i+m_{i'}}{m}  
\sum_{j'} s_{j'} \beta_{j'} C^-_{j'} \frac{m_j+m_{j'}}{m}|G^+_{i'}\rangle |G^-_{j'}\rangle \\
&&- \frac{g_0^2}{2}\beta_i \beta_j  Y^-(\vec{R})  \sum_{j'} s_{j'}\beta_{j'} C^-_{j'}
\frac{m_j+m_{j'}}{m}
\sum_{i'} s_{i'} \beta_{i'} C^+_{i'} \frac{m_i+m_{i'}}{m}|G^+_{i'}\rangle|G^-_{j'}\rangle \nonumber \\
&&=E C^+_iC^-_j|G^+_i\rangle|G^-_j\rangle. \nonumber
\eea
According to (\ref{eq:EVP1}), with $E^\pm=m$, the first curly bracket is simply 
$mC^+_i|G^+_i\rangle$, and the second is $mC^-_j|G^-_j\rangle$.  These two terms
contribute $2m C^+_i C^-_j|G^+_i\rangle|G^-_j\rangle$ to the equation and, when
subtracted from both sides, leave the effective potential $V_{\rm eff}(\vec{R})\equiv E-2m$ 
determined by 
\bea  \label{eq:Veff}
&-& \frac{g_0^2}{2} \beta_i \beta_j Y(\vec{R}) \sum_{i'} s_{i'}\beta_{i'}  C^+_{i'} 
\frac{m_i+m_{i'}}{m}  
\sum_{j'} s_{j'} \beta_{j'} C^-_{j'} \frac{m_j+m_{j'}}{m}|G^+_{i'}\rangle |G^-_{j'}\rangle \\
&&=V_{\rm eff} C^+_iC^-_j|G^+_i\rangle|G^-_j\rangle, \nonumber
\eea
with $Y\equiv Y^++Y^-$.

To extract $V_{\rm eff}$, we define $|G^\pm\rangle\equiv\sum_i C^\pm_is_i\beta_i|G^\pm_i\rangle$,
multiply Eq.~(\ref{eq:Veff}) by $s_i\beta_i s_j\beta_j$, and sum over $i$ and $j$
\bea
\lefteqn{-\frac{g_0^2}{2} Y(\vec{R}) \sum_i s_i \beta_i^2 \sum_j s_j \beta_j^2 
\left[\frac{m_i}{m}|G^+\rangle+\sum_{i'} s_{i'}\beta_{i'}  C^+_{i'}\frac{m_{i'}}{m} |G^+_{i'}\rangle\right]}&& \\
&& \rule{1in}{0mm}  \times \left[\frac{m_j}{m}|G^-\rangle+ \sum_{j'} s_{j'} \beta_{j'} C^-_{j'} \frac{m_{j'}}{m} |G^-_{j'}\rangle \right]
=V_{\rm eff} |G^+\rangle|G^-\rangle, \nonumber
\eea
To simplify the result further, we use the constraint $\sum_is_i\beta_i^2=0$, which eliminates
all terms between the square brackets except for the product of the first in each.  We can then
equate coefficients of $|G^+\rangle|G^-\rangle$ to obtain
\be
V_{\rm eff}=-\frac{g_0^2}{2} Y(\vec{R}) \left(\sum_i \frac{s_i\beta_i^2 m_i}{m}\right)^2.
\ee
The integrals in $Y$ are computed in \cite{ScalarYukawa}.  They yield
\be
Y(\vec{R})=\sum_k r_k \xi_k^2 \frac{e^{-\mu_k R}}{8\pi R}.
\ee
In the limit of infinite PV boson masses, with $r_0=1$,
$\xi_0=1$, and $\mu_0=\mu$, this reduces to $e^{-\mu R}/(8 \pi R)$,
and the effective potential is found to be
\be
V_{\rm eff}=-\frac{g_0^2}{2}  \left(\sum_i \frac{s_i\beta_i^2 m_i}{m}\right)^2 \frac{e^{-\mu R}}{8 \pi R}.
\ee
This is the expression given in (\ref{eq:VeffFinal}).


\end{document}